\documentclass[12pt]{article}
\usepackage{amsfonts}
\usepackage{amsmath}
\usepackage{amssymb}

\input{tcilatex}
\begin{document}

\begin{center}
\smallskip \ 

\textbf{DIRAC EQUATION IN (1+3)}

\smallskip \ 

\textbf{AND (2+2) DIMENSIONS}

\smallskip \ 

J. A. Nieto \footnote{%
nieto@uas.edu.mx, janieto1@asu.edu} and C. Pereyra \footnote{%
pereyra.claudio@gmail.com},

\smallskip \ 

\smallskip \ 

\smallskip

\textit{Facultad de Ciencias F\'{\i}sico-Matem\'{a}ticas de la Universidad
Aut\'{o}noma} \textit{de Sinaloa, 80010, Culiac\'{a}n Sinaloa, M\'{e}xico.}

\smallskip

\smallskip

\textbf{Abstract}
\end{center}

We develop a systematic method to derive the Majorana representation of the
Dirac equation in (1+3)-dimensions. We compare with similar approach in
(2+2)-dimensions . We argue that our formalism can be useful to have a
better understanding of possible Majorana fermions.

\bigskip \ 

\bigskip \ 

\bigskip \ 

\bigskip \ 

\bigskip \ 

\bigskip \ 

\bigskip \ 

\bigskip \ 

\bigskip \ 

Keywords: Dirac equation, (2+2)-dimensions, Majorana representation.

Pacs numbers: 04.20.Gz, 04.60.-Ds, 11.30.Ly

July 18, 2013

\newpage

\noindent \textbf{1. Introduction}

\smallskip \ 

\noindent It is well known that the Dirac equation [1] admits a physical
formalism in which both the gamma matrices $\gamma ^{\mu }$ and the physical
states (spinors) $\psi $ are in the so called Majorana representation $%
\gamma _{M}^{\mu }$ and $\psi _{M}$ (Here, the $M$ in $\gamma _{M}^{\mu }$
and $\psi _{M}$ stands for Majorana.) It turns out that, in this case, all
components in $\gamma _{M}^{\mu }$ can be chosen imaginary. When one imposes
the additional condition $\psi _{M}^{c}=\psi _{M}$, where the $c$ in $\psi
_{M}^{c}$ means charge conjugation, then $\psi _{M}$ becomes a real spinor,
that is $\psi _{M}=\psi _{M}^{\ast }$. This condition defines the so called
Majorana fermions which are of great interest in various scenarios,
including superstrings (see Ref. [2] and references therein) and dark matter
theories (see Ref. [3] and references therein).

The above analysis refers to the Dirac equation in (1+3)-dimensions and one
wonders what could be the corresponding results for the Dirac equation in
(2+2)-dimensions. Of course even from the beginning one knows that in
(2+2)-dimensions there exist Majorana and Weyl spinors, while in the case of
(1+3)-dimensions the spinors can be either Majorana or Weyl, but no both.
This is an important difference and one would like to understand this
phenomenon deeply. There are general analysis about the existence of
Majorana and Weyl spinors in various dimensions and signatures (see Ref.
[4]), but it will interesting if such differences, between (1+3) and (2+2)
dimensions, are understood from first principles. For this purpose, we first
develop a method for constructing the Dirac equation in (1+3)-dimensions in
the three typical context, namely Weyl, Dirac and Majorana representations.
We then apply a similar method to the case of the Dirac equation in
(2+2)-dimensions.

Before we address the problem at hand let us just mention some physical
reasons why one may be interested in the Dirac equation in (2+2)-dimensions.
First of all, it has been proved that the (2+2)-signature can be considered
as exceptional signature [5]. In fact, from the perspective of 2t-physics
[6] the structure of (2+2)-physics can be considered as a minimal model.
Secondly, it is known that the (2+2)-signature is deeply linked to the $%
SL(2,R)$-group, which is very important structure in string theory, two
dimensional black holes [7] and conformal field theory [8]. Moreover, the
(2+2)-signature emerges as an important physical concept in a number of
physical scenarios, including background for $N=2$ strings [9]-[10] (see
also Refs [11]-[14]), Yang-Mills theory in Atiyah-Singer background [15]
(see also Refs. [16] for the importance of the (2+2)-signature in
mathematics), Majorana-Weyl spinor in supergravity [13] and more recently in
loop quantum gravity in terms of oriented matroid theory [17].

The structure of this paper is as follows. In section 2, we briefly review
the Dirac's method of constraint Hamiltonian systems. In sections 3, we
develop a straightforward method of writing the Dirac equation in the Weyl,
Dirac and Majorana representations. In section 4, we apply similar method to
the case of the Dirac equation in (2+2)-dimensions, remarking in this case
the important role played by the $SL(2,R)$-group. In section 5, we make some
additional comments.

\bigskip \ 

\noindent \textbf{2. Lagrange-Hamiltonian formalism for constrainted systems}

\smallskip \ 

\noindent Let us consider the action

\begin{equation}
S[q(t)]=\int dtL(q,\dot{q}),  \tag{1}
\end{equation}%
where the Lagrangian $L=L(q,\dot{q})$ is a function of the $q^{i}$%
-coordinates and the corresponding velocities $\dot{q}^{i}\equiv dq^{i}/dt,$
with $i,j=1,\dots ,n$.

The canonical momentum $p_{i}$ conjugate to $q^{i}$ is defined by

\begin{equation}
p_{i}\equiv \frac{\partial L}{\partial \dot{q}^{i}}.  \tag{2}
\end{equation}%
Thus, the action (1) can be rewritten in the alternative form

\begin{equation}
S[q(t),p(t)]=\int dt(\dot{q}^{i}p_{i}-H_{c}),  \tag{3}
\end{equation}%
where $H_{c}=H_{c}(q,p)$ is the canonical Hamiltonian,

\begin{equation}
H_{c}(q,p)\equiv \dot{q}^{i}p_{i}-L.  \tag{4}
\end{equation}

If one considers $m$ first class Hamiltonian constraints $H_{A}(q,p)\approx
0 $, (here the symbol "$\approx $" means weakly equal to zero [18]), with $%
A=1,2...,m$, then the action (3) can be generalized as follows:

\begin{equation}
S[q,p]=\int dt(\dot{q}^{i}p_{i}-H_{c}-\lambda ^{A}H_{A}).  \tag{5}
\end{equation}%
Here, the symbol $\lambda ^{A}$ denotes arbitrary Lagrange multiplier.

The Poisson bracket for arbitrary functions $f(q,p)$ and $g(q,p)$ of the
canonical variables $q^{i}$ and $p_{i}$ is defined as usual,

\begin{equation}
\{f,g\}=\frac{\partial f}{\partial q^{i}}\frac{\partial g}{\partial p_{i}}-%
\frac{\partial f}{\partial p_{i}}\frac{\partial g}{\partial q^{i}}.  \tag{6}
\end{equation}%
Using (6) we find that

\begin{equation}
\begin{array}{c}
\{q^{i},q^{j}\}=0, \\ 
\\ 
\{q^{i},p_{j}\}=\delta _{j}^{i}, \\ 
\\ 
\{p_{i},p_{j}\}=0.%
\end{array}
\tag{7}
\end{equation}%
The symbol $\delta _{j}^{i}$ denotes a Kronecker delta.

Let us apply the above method to the case of a relativistic point particle.
Assume that the motion of a relativistic point particle is describe by the
coordinates $x^{\mu }(\tau )$, where $\tau $ is an arbitrary parameter. The
dynamics of this system is obtained from the Lagrangian

\begin{equation}
L=-m_{0}(-\dot{x}^{\mu }\dot{x}^{\nu }\eta _{\mu \nu })^{1/2},  \tag{8}
\end{equation}%
where $\eta _{\mu \nu }$ is a flat metric. Using (8) one finds that the
canonical momentum

\begin{equation}
p_{\mu }=\frac{\partial L}{\partial \dot{x}^{\mu }},  \tag{9}
\end{equation}%
becomes

\begin{equation}
p_{\mu }=-\frac{m_{0}\eta _{\mu \nu }\dot{x}^{\nu }}{(-\dot{x}^{\alpha }\dot{%
x}^{\beta }\eta _{\alpha \beta })^{1/2}}.  \tag{10}
\end{equation}%
Thus, from (10) one obtains the identity

\begin{equation}
p^{\mu }p^{\nu }\eta _{\mu \nu }+m_{0}^{2}=0,  \tag{11}
\end{equation}%
which is, indeed, a first class constraint.

It turns out that by promoting $p_{\mu }$ as an operator $\hat{p}_{\mu }=-i%
\frac{\partial }{\partial x^{\mu }}$, the constraint (11) leads to the
Klein-Gordon equation,%
\begin{equation}
(\eta ^{\mu \nu }\hat{p}_{\mu }\hat{p}_{\nu }+m_{0}^{2})\psi =0.  \tag{12}
\end{equation}%
In this case one finds that the physical states $\psi $ may be associated
with either spin-$0$ or spin-$\frac{1}{2}$ particle (see Ref. [18] for
details).

\bigskip \ 

\bigskip \ 

\bigskip \ 

\bigskip \ 

\noindent \textbf{3. Dirac equation in (1+3)-dimensions}

\smallskip \ 

In order to derive the Dirac equation in (1+3)-dimensions, by starting from
(12), we first introduce the Pauli matrices,

\begin{equation}
\sigma ^{1}=\left( 
\begin{array}{cc}
0 & 1 \\ 
1 & 0%
\end{array}%
\right) \ ,\  \sigma ^{2}=\left( 
\begin{array}{cc}
0 & -i \\ 
i & 0%
\end{array}%
\right) \ ,\  \sigma ^{3}=\left( 
\begin{array}{cc}
1 & 0 \\ 
0 & -1%
\end{array}%
\right) .  \tag{13}
\end{equation}%
It is not difficult to see that $\sigma ^{i}$, with $i$ running from $1$ to $%
3$, satisfies

\begin{equation}
\sigma ^{i}\sigma ^{j}+\sigma ^{j}\sigma ^{i}=2\delta ^{ij}  \tag{14}
\end{equation}%
and

\begin{equation}
\sigma ^{i}\sigma ^{j}-\sigma ^{j}\sigma ^{i}=2i\varepsilon ^{ijk}\sigma
_{k}.  \tag{15}
\end{equation}%
Here, $\delta ^{ij}=diag(1,1,1)$ is again the Kronecker delta and $%
\varepsilon ^{ijk}$ is the completely antisymmetric $\varepsilon $-symbol,
with $\varepsilon ^{123}=1$. Moreover, by adding (14) and (15) one obtains a
useful formula,%
\begin{equation}
\sigma ^{i}\sigma ^{j}=\delta ^{ij}+i\varepsilon ^{ijk}\sigma _{k}.  \tag{16}
\end{equation}

In (1+3)-dimensions we have $\eta ^{\mu \nu }=diag(-1,1,1,1)$ and therefore
(12) becomes%
\begin{equation}
(-\hat{p}_{0}\hat{p}_{0}+\delta ^{ij}\hat{p}_{i}\hat{p}_{j}+m_{0}^{2})\psi
=0.  \tag{17}
\end{equation}%
Using (14) one discovers that (17) can be written as

\begin{equation}
(-\hat{p}_{0}\hat{p}_{0}+\sigma ^{i}\sigma ^{j}\hat{p}_{i}\hat{p}%
_{j}+m_{0}^{2})\psi =0.  \tag{18}
\end{equation}%
In turn this expression can also be expressed as

\begin{equation}
(-\hat{p}_{0}+\sigma ^{i}\hat{p}_{i})(\hat{p}_{0}+\sigma ^{j}\hat{p}%
_{j})\psi +m_{0}^{2}\psi =0.  \tag{19}
\end{equation}%
It is convenient to rename $\psi $ in the form $\psi _{L}\equiv \psi $ and
to introduce the definition

\begin{equation}
\psi _{R}\equiv -\frac{1}{m_{0}}(\hat{p}_{0}+\sigma ^{j}\hat{p}_{j})\psi
_{L}.  \tag{20}
\end{equation}%
So (19) yields

\begin{equation}
-(-\hat{p}_{0}+\sigma ^{i}\hat{p}_{i})\psi _{R}+m_{0}\psi _{L}=0.  \tag{21}
\end{equation}%
Let us write (20) and (21) in the form

\begin{equation}
(\hat{p}_{0}+\sigma ^{j}\hat{p}_{j})\psi _{L}+m_{0}\psi _{R}=0  \tag{22}
\end{equation}%
and

\begin{equation}
(\hat{p}_{0}-\sigma ^{i}\hat{p}_{i})\psi _{R}+m_{0}\psi _{L}=0,  \tag{23}
\end{equation}%
respectively. One can put together the equations (22) and (23);

\begin{equation}
\left[ \left( 
\begin{array}{cc}
0 & 1 \\ 
1 & 0%
\end{array}%
\right) \hat{p}_{0}+\left( 
\begin{array}{cc}
0 & \sigma ^{i} \\ 
-\sigma ^{i} & 0%
\end{array}%
\right) \hat{p}_{i}+\left( 
\begin{array}{cc}
1 & 0 \\ 
0 & 1%
\end{array}%
\right) m_{0}\right] \left( 
\begin{array}{c}
\psi _{R} \\ 
\psi _{L}%
\end{array}%
\right) =0.  \tag{24}
\end{equation}%
We recognize in (24) the Dirac equation in the Weyl representation,

\begin{equation}
\left[ \gamma _{W}^{0}\hat{p}_{0}+\gamma _{W}^{i}\hat{p}_{i}+m_{0}\right]
\psi _{W}=0,  \tag{25}
\end{equation}%
which in covariant notation becomes

\begin{equation}
\left[ \gamma _{W}^{\mu }\hat{p}_{\mu }+m_{0}\right] \psi _{W}=0,  \tag{26}
\end{equation}%
where%
\begin{equation}
\gamma _{W}^{0}=\left( 
\begin{array}{cc}
0 & 1 \\ 
1 & 0%
\end{array}%
\right) \  \ ,\  \  \gamma _{W}^{i}=\left( 
\begin{array}{cc}
0 & \sigma ^{i} \\ 
-\sigma ^{i} & 0%
\end{array}%
\right) ,  \tag{27}
\end{equation}%
and

\begin{equation}
\psi _{W}=\left( 
\begin{array}{c}
\psi _{R} \\ 
\psi _{L}%
\end{array}%
\right) .  \tag{28}
\end{equation}%
Here, the $W$ in $\gamma _{W}^{\mu }$ and $\psi _{W}$ means that this
quantities are in the Weyl representation. One can verify that the
expressions

\begin{equation*}
\left( \gamma _{W}^{0}\right) ^{2}=I\text{ \  \ , \  \ }\left( \gamma
_{W}^{1}\right) ^{2}=\left( \gamma _{W}^{1}\right) ^{2}=\left( \gamma
_{W}^{1}\right) ^{2}=-1,
\end{equation*}%
can be derived from%
\begin{equation}
\gamma _{W}^{\mu }\gamma _{W}^{\nu }+\gamma _{W}^{\nu }\gamma _{W}^{\mu
}=-2\eta ^{\mu \nu }.  \tag{29}
\end{equation}

Now, by adding and subtracting (22) and (23) one finds

\begin{equation}
\hat{p}_{0}(\psi _{R}+\psi _{L})+\sigma ^{i}\hat{p}_{i}(\psi _{L}-\psi
_{R})+m_{0}(\psi _{R}+\psi _{L})=0  \tag{30}
\end{equation}%
and%
\begin{equation}
-\hat{p}_{0}(\psi _{L}-\psi _{R})-\sigma ^{i}\hat{p}_{i}(\psi _{R}+\psi
_{L})+m_{0}(\psi _{L}-\psi _{R})=0,  \tag{31}
\end{equation}%
respectively. So, by defining

\begin{equation}
\begin{array}{c}
\psi _{A}=\psi _{R}+\psi _{L}, \\ 
\\ 
\psi _{B}=\psi _{L}-\psi _{R},%
\end{array}
\tag{32}
\end{equation}%
one gets

\begin{equation}
\hat{p}_{0}\psi _{A}+\sigma ^{i}\hat{p}_{i}\psi _{B}+m_{0}\psi _{A}=0, 
\tag{33}
\end{equation}%
and

\begin{equation}
-\hat{p}_{0}\psi _{B}-\sigma ^{i}\hat{p}_{i}\psi _{A}+m_{0}\psi _{B}=0. 
\tag{34}
\end{equation}%
Thus, (33) and (34) lead to

\begin{equation}
\left[ \left( 
\begin{array}{cc}
1 & 0 \\ 
0 & -1%
\end{array}%
\right) \hat{p}_{0}+\left( 
\begin{array}{cc}
0 & \sigma ^{i} \\ 
-\sigma ^{i} & 0%
\end{array}%
\right) \hat{p}_{i}+\left( 
\begin{array}{cc}
1 & 0 \\ 
0 & 1%
\end{array}%
\right) m_{0}\right] \left( 
\begin{array}{c}
\psi _{A} \\ 
\psi _{B}%
\end{array}%
\right) =0,  \tag{35}
\end{equation}%
and consequently using the notation

\begin{equation}
\gamma _{D}^{0}=\left( 
\begin{array}{cc}
1 & 0 \\ 
0 & -1%
\end{array}%
\right) \text{ \  \ , }\  \  \gamma _{D}^{i}=\left( 
\begin{array}{cc}
0 & \sigma ^{i} \\ 
-\sigma ^{i} & 0%
\end{array}%
\right) ,  \tag{36}
\end{equation}%
the expressions (33) and (34) become%
\begin{equation}
\left[ \gamma _{D}^{0}\hat{p}_{0}+\gamma _{D}^{i}\hat{p}_{i}+m_{0}\right]
\psi _{D}=0,  \tag{37}
\end{equation}%
or

\begin{equation}
\left[ \gamma _{D}^{\mu }\hat{p}_{\mu }+m_{0}\right] \psi _{D}=0.  \tag{38}
\end{equation}%
Here the $D$ in $\gamma _{D}^{\mu }$ and $\psi _{D}$ stands for Dirac
representation.

The Weyl and the Dirac representations are related by

\begin{equation}
\left( 
\begin{array}{c}
\psi _{A} \\ 
\psi _{B}%
\end{array}%
\right) =\left( 
\begin{array}{cc}
1 & 1 \\ 
-1 & 1%
\end{array}%
\right) \left( 
\begin{array}{c}
\psi _{R} \\ 
\psi _{L}%
\end{array}%
\right) =\left( 
\begin{array}{c}
\psi _{R}+\psi _{L} \\ 
\psi _{L}-\psi _{R}%
\end{array}%
\right) .  \tag{39}
\end{equation}%
By defining

\begin{equation}
W=\left( 
\begin{array}{cc}
1 & 1 \\ 
-1 & 1%
\end{array}%
\right) ,  \tag{40}
\end{equation}%
the expression (39) means $\psi _{D}=W\psi _{W}$. One also finds,

\begin{equation}
W^{-1}=\frac{1}{2}\left( 
\begin{array}{cc}
1 & -1 \\ 
1 & 1%
\end{array}%
\right) .  \tag{41}
\end{equation}%
Thus, it is not difficult to see that $\gamma _{W}^{\mu }$ and $\gamma
_{D}^{\mu }$ are related by

\begin{equation}
\gamma _{D}^{\mu }=W\gamma _{W}^{\mu }W^{-1}.  \tag{42}
\end{equation}

Now, let us expand (33) and (34) in the form

\begin{equation}
\hat{p}_{0}\psi _{A}+\sigma ^{1}\hat{p}_{1}\psi _{B}+\sigma ^{2}\hat{p}%
_{2}\psi _{B}+\sigma ^{3}\hat{p}_{3}\psi _{B}+m_{0}\psi _{A}=0,  \tag{43}
\end{equation}%
and

\begin{equation}
-\hat{p}_{0}\psi _{B}-\sigma ^{1}\hat{p}_{1}\psi _{A}-\sigma ^{2}\hat{p}%
_{2}\psi _{A}-\sigma ^{3}\hat{p}_{3}\psi _{A}+m_{0}\psi _{B}=0.  \tag{44}
\end{equation}%
Multiplying (43) and (44) by $\sigma ^{2}$ one finds

\begin{equation}
\hat{p}_{0}\sigma ^{2}\psi _{A}+\sigma ^{2}\sigma ^{1}\hat{p}_{1}\psi
_{B}+\sigma ^{2}\sigma ^{2}\hat{p}_{2}\psi _{B}+\sigma ^{2}\sigma ^{3}\hat{p}%
_{3}\psi _{B}+m_{0}\sigma ^{2}\psi _{A}=0  \tag{45}
\end{equation}%
and

\begin{equation}
-\hat{p}_{0}\sigma ^{2}\psi _{B}-\sigma ^{2}\sigma ^{1}\hat{p}_{1}\psi
_{A}-\sigma ^{2}\sigma ^{2}\hat{p}_{2}\psi _{A}-\sigma ^{2}\sigma ^{3}\hat{p}%
_{3}\psi _{A}+m_{0}\sigma ^{2}\psi _{B}=0,  \tag{46}
\end{equation}%
respectively. Since from (16) one knows that

\begin{equation}
\begin{array}{c}
\sigma ^{1}\sigma ^{2}=i\sigma ^{3}\  \ ,\  \  \sigma ^{2}\sigma ^{1}=-i\sigma
^{3}, \\ 
\\ 
\sigma ^{2}\sigma ^{3}=i\sigma ^{1}\text{ \  \ , \  \ }\sigma ^{3}\sigma
^{2}=-i\sigma ^{1},%
\end{array}
\tag{47}
\end{equation}%
by adding (43) and (46) one sees that

\begin{equation}
\begin{array}{c}
-\sigma ^{2}\hat{p}_{0}(\psi _{B}-\sigma ^{2}\psi _{A})+i\sigma ^{3}\hat{p}%
_{1}(\psi _{A}+\sigma ^{2}\psi _{B})+\sigma ^{2}\hat{p}_{2}(\psi _{B}-\sigma
^{2}\psi _{A}) \\ 
\\ 
-i\sigma ^{1}\hat{p}_{3}(\psi _{A}+\sigma ^{2}\psi _{B})+m_{0}(\psi
_{A}+\sigma ^{2}\psi _{B})=0,%
\end{array}
\tag{48}
\end{equation}%
while subtracting (44) and (45) one gets

\begin{equation}
\begin{array}{c}
-\sigma ^{2}\hat{p}_{0}(\psi _{A}+\sigma ^{2}\psi _{B})+i\sigma ^{3}\hat{p}%
_{1}(\psi _{B}-\sigma ^{2}\psi _{A})-\sigma ^{2}\hat{p}_{2}(\psi _{A}+\sigma
^{2}\psi _{B}) \\ 
\\ 
-i\sigma ^{1}\hat{p}_{3}(\psi _{B}-\sigma ^{2}\psi _{A})+m_{0}(\psi
_{B}-\sigma ^{2}\psi _{A})=0.%
\end{array}
\tag{49}
\end{equation}%
Defining%
\begin{equation}
\begin{array}{c}
\psi _{1}=\psi _{A}+\sigma ^{2}\psi _{B} \\ 
\\ 
\psi _{2}=\psi _{B}-\sigma ^{2}\psi _{A},%
\end{array}
\tag{50}
\end{equation}%
one finds that the expressions (48) and (49) become

\begin{equation}
-\sigma ^{2}\hat{p}_{0}\psi _{2}+i\sigma ^{3}\hat{p}_{1}\psi _{1}+\sigma ^{2}%
\hat{p}_{2}\psi _{2}-i\sigma ^{1}\hat{p}_{3}\psi _{1}+m\psi _{1}=0  \tag{51}
\end{equation}%
and%
\begin{equation}
-\sigma ^{2}\hat{p}_{0}\psi _{1}+i\sigma ^{3}\hat{p}_{1}\psi _{2}-\sigma ^{2}%
\hat{p}_{2}\psi _{1}-i\sigma ^{1}\hat{p}_{3}\psi _{2}+m_{0}\psi _{2}=0, 
\tag{52}
\end{equation}%
respectively. In turn the equations (51) and (52) can be written as

\begin{equation}
\begin{array}{c}
\left[ 
\begin{array}{c}
\left( 
\begin{array}{cc}
0 & -\sigma ^{2} \\ 
-\sigma ^{2} & 0%
\end{array}%
\right) \hat{p}_{0}+\left( 
\begin{array}{cc}
i\sigma ^{3} & 0 \\ 
0 & i\sigma ^{3}%
\end{array}%
\right) \hat{p}_{1}+\left( 
\begin{array}{cc}
0 & \sigma ^{2} \\ 
-\sigma ^{2} & 0%
\end{array}%
\right) \hat{p}_{2} \\ 
\\ 
+\left( 
\begin{array}{cc}
-i\sigma ^{1} & 0 \\ 
0 & -i\sigma ^{1}%
\end{array}%
\right) \hat{p}_{3}+\left( 
\begin{array}{cc}
1 & 0 \\ 
0 & 1%
\end{array}%
\right) m_{0}%
\end{array}%
\right] \\ 
\\ 
\times \left( 
\begin{array}{c}
\psi _{1} \\ 
\psi _{2}%
\end{array}%
\right) =0.%
\end{array}
\tag{53}
\end{equation}%
Hence, by using the definitions%
\begin{equation}
\begin{array}{c}
\gamma _{M}^{0}=\left( 
\begin{array}{cc}
0 & -\sigma ^{2} \\ 
-\sigma ^{2} & 0%
\end{array}%
\right) ,\text{ \ }\gamma _{M}^{1}=\left( 
\begin{array}{cc}
i\sigma ^{3} & 0 \\ 
0 & i\sigma ^{3}%
\end{array}%
\right) \text{, \ }\gamma _{M}^{2}=\left( 
\begin{array}{cc}
0 & \sigma ^{2} \\ 
-\sigma ^{2} & 0%
\end{array}%
\right) \\ 
\\ 
\text{\  \  \  \ }\gamma _{M}^{3}=\left( 
\begin{array}{cc}
i\sigma ^{1} & 0 \\ 
0 & i\sigma ^{1}%
\end{array}%
\right) \text{, \ }\psi _{M}=\left( 
\begin{array}{c}
\psi _{A}+\sigma ^{2}\psi _{B} \\ 
\psi _{B}-\sigma ^{2}\psi _{A}%
\end{array}%
\right) ,%
\end{array}
\tag{54}
\end{equation}%
one observes that the equation (53) can be rewritten as%
\begin{equation}
\left[ \gamma _{M}^{0}\hat{p}_{0}+\gamma _{M}^{1}\hat{p}_{1}+\gamma _{M}^{2}%
\hat{p}_{2}+\gamma _{M}^{3}\hat{p}_{3}+m_{0}\right] \text{\ }\psi _{M}=0 
\tag{55}
\end{equation}%
or

\begin{equation}
\left[ \gamma _{M}^{\mu }\hat{p}_{\mu }+m_{0}\right] \text{\ }\psi _{M}=0. 
\tag{56}
\end{equation}%
This is the Dirac equation in the Majorana representation. One can verify
that

\begin{equation}
\gamma _{M}^{\mu }\gamma _{M}^{\nu }+\gamma _{M}^{\nu }\gamma _{M}^{\mu
}=-2\eta ^{\mu \nu }.  \tag{57}
\end{equation}

One can link the Dirac and the Majorana representations by considering the
matrix

\begin{equation}
V=\left( 
\begin{array}{cc}
1 & \sigma ^{2} \\ 
-\sigma ^{2} & 1%
\end{array}%
\right) ,  \tag{58}
\end{equation}%
and its inverse%
\begin{equation}
V^{-1}=\frac{1}{2}\left( 
\begin{array}{cc}
1 & -\sigma ^{2} \\ 
\sigma ^{2} & 1%
\end{array}%
\right) .  \tag{59}
\end{equation}%
In fact, one has

\begin{equation}
\left( 
\begin{array}{c}
\psi _{1} \\ 
\psi _{2}%
\end{array}%
\right) =\left( 
\begin{array}{cc}
1 & \sigma ^{2} \\ 
-\sigma ^{2} & 1%
\end{array}%
\right) \left( 
\begin{array}{c}
\psi _{A} \\ 
\psi _{B}%
\end{array}%
\right) =\left( 
\begin{array}{c}
\psi _{A}+\sigma ^{2}\psi _{B} \\ 
\psi _{B}-\sigma ^{2}\psi _{A}%
\end{array}%
\right)  \tag{60}
\end{equation}%
and

\begin{equation}
\gamma _{M}^{\mu }=V\gamma _{D}^{\mu }V^{-1}.  \tag{61}
\end{equation}

It is interesting to observe that since $\gamma _{D}^{\mu }=W\gamma
_{W}^{\mu }W^{-1}$, one obtains that

\begin{equation}
\gamma _{M}^{\mu }=VW\gamma _{W}^{\mu }W^{-1}V^{-1}.  \tag{62}
\end{equation}%
This means that by using the matrix

\begin{equation}
\begin{array}{c}
U=VW=\left( 
\begin{array}{cc}
1 & \sigma ^{2} \\ 
-\sigma ^{2} & 1%
\end{array}%
\right) \left( 
\begin{array}{cc}
1 & 1 \\ 
-1 & 1%
\end{array}%
\right) \\ 
\\ 
\text{ \  \  \  \ }=\left( 
\begin{array}{cc}
\left( 1-\sigma ^{2}\right) & \left( 1+\sigma ^{2}\right) \\ 
-\left( \sigma ^{2}+1\right) & \left( 1-\sigma ^{2}\right)%
\end{array}%
\right) ,%
\end{array}
\tag{63}
\end{equation}%
and its inverse

\begin{equation}
\begin{array}{c}
U^{-1}=W^{-1}V^{-1}=\frac{1}{4}\left( 
\begin{array}{cc}
1 & -1 \\ 
1 & 1%
\end{array}%
\right) \left( 
\begin{array}{cc}
1 & -\sigma ^{2} \\ 
\sigma ^{2} & 1%
\end{array}%
\right) \\ 
\\ 
\text{ \  \  \  \  \  \  \ }=\frac{1}{4}\left( 
\begin{array}{cc}
\left( 1-\sigma ^{2}\right) & -\left( \sigma ^{2}+1\right) \\ 
\left( 1+\sigma ^{2}\right) & \left( 1-\sigma ^{2}\right)%
\end{array}%
\right) ,%
\end{array}
\tag{64}
\end{equation}%
one can transform the Weyl and the Majorana $\gamma ^{\mu }$ matrices. In
fact one can verify that%
\begin{equation}
\gamma _{M}^{\mu }=\frac{1}{4}\left( 
\begin{array}{cc}
\left( 1-\sigma ^{2}\right) & \left( 1+\sigma ^{2}\right) \\ 
-\left( \sigma ^{2}+1\right) & \left( 1-\sigma ^{2}\right)%
\end{array}%
\right) \gamma _{W}^{\mu }\left( 
\begin{array}{cc}
\left( 1-\sigma ^{2}\right) & -\left( \sigma ^{2}+1\right) \\ 
\left( 1+\sigma ^{2}\right) & \left( 1-\sigma ^{2}\right)%
\end{array}%
\right) .  \tag{65}
\end{equation}%
Similarly, one discovers that the spinors $\psi _{M}$ and $\psi _{W}$ are
related by $\psi _{M}=U\psi _{W}$.

\bigskip \ 

\bigskip \ 

\noindent \textbf{4.- The Dirac equation in (2+2)-dimensions}

\smallskip \ 

We shall start by assuming that the flat metric $\eta _{\mu \nu }$ has now
the form%
\begin{equation}
\eta _{\mu \nu }\longrightarrow g_{\mu \nu }=diag(-1,1,-1,1).  \tag{66}
\end{equation}%
In this case the Pauli matrices $\sigma ^{i}$ will replaced by $\rho ^{i}$,
where%
\begin{equation}
\begin{array}{ccc}
\rho ^{1}=%
\begin{pmatrix}
0 & 1 \\ 
1 & 0%
\end{pmatrix}%
, & \rho ^{2}=%
\begin{pmatrix}
0 & -1 \\ 
1 & 0%
\end{pmatrix}%
, & \rho ^{3}=%
\begin{pmatrix}
1 & 0 \\ 
0 & -1%
\end{pmatrix}%
.%
\end{array}
\tag{67}
\end{equation}%
Observe that the $\rho ^{i}$ satisfies

\begin{equation}
\rho ^{i}\rho ^{j}+\rho ^{j}\rho ^{i}=2g^{ij}  \tag{68}
\end{equation}%
and

\begin{equation}
\rho ^{i}\rho ^{j}-\rho ^{j}\rho ^{i}=-2\epsilon ^{ijk}g_{kl}\rho ^{l}, 
\tag{69}
\end{equation}%
where $g_{ij}=diag(1,-1,1)$ and $\epsilon ^{ijk}$ is a completely
antisymmetric $\epsilon $-symbol with $\epsilon ^{123}=1$ and $\epsilon
_{123}=-1.$

Again one first considers the quantum equation

\begin{equation}
\lbrack g^{\mu \nu }\hat{p}_{\mu }\hat{p}_{\nu }+M_{0}^{2}]\varphi =0, 
\tag{70}
\end{equation}%
which can also be written as%
\begin{equation}
\lbrack -\hat{p}_{0}\hat{p}_{0}+g^{ij}\hat{p}_{i}\hat{p}_{j}+M_{0}^{2}]%
\varphi =0.  \tag{71}
\end{equation}%
Here, $M_{0}$ is the analogue in (2+2)-dimensions of the rest mass $m_{0}$
in (1+3)-dimensions. Using (68) one sees that (69) can be written as%
\begin{equation}
\lbrack (-\hat{p}_{0}+\rho ^{i}\hat{p}_{i})(\hat{p}_{0}+\rho ^{j}\hat{p}%
_{j})+M_{0}^{2}]\varphi =0.  \tag{72}
\end{equation}

Now, we define two spinors

\begin{equation}
\varphi _{L}\equiv \varphi  \tag{73}
\end{equation}%
and

\begin{equation}
\varphi _{R}\equiv -\frac{1}{M_{0}}(\hat{p}_{0}+\rho ^{i}\hat{p}_{i})\varphi
_{L}.  \tag{74}
\end{equation}%
Explicitly (74) leads to%
\begin{equation}
(\hat{p}_{0}+\rho ^{i}\hat{p}_{i})\varphi _{L}+M_{0}\varphi _{R}=0,  \tag{75}
\end{equation}%
while (72) and (74) give%
\begin{equation}
(\hat{p}_{0}-\rho ^{i}\hat{p}_{i})\varphi _{R}+M_{0}\varphi _{L}=0.  \tag{76}
\end{equation}%
These two equations can be expressed in a matrix form%
\begin{equation}
\left( 
\begin{bmatrix}
0 & 1 \\ 
1 & 0%
\end{bmatrix}%
\hat{p}_{0}+%
\begin{bmatrix}
0 & \rho ^{i} \\ 
-\rho ^{i} & 0%
\end{bmatrix}%
\hat{p}_{i}+%
\begin{bmatrix}
1 & 0 \\ 
0 & 1%
\end{bmatrix}%
M_{0}\right) \left( 
\begin{array}{c}
\varphi _{R} \\ 
\varphi _{L}%
\end{array}%
\right) =0.  \tag{77}
\end{equation}%
Of course, one can write (77) in a more compact form%
\begin{equation}
(\Gamma _{W}^{\mu }\hat{p}_{\mu }+M_{0})\varphi _{W}=0.  \tag{78}
\end{equation}%
Here, we used the following definitions;

\begin{equation}
\varphi _{W}\equiv \left( 
\begin{array}{c}
\varphi _{R} \\ 
\varphi _{L}%
\end{array}%
\right) ,  \tag{79}
\end{equation}%
\begin{equation}
\Gamma _{W}^{0}\equiv 
\begin{bmatrix}
0 & 1 \\ 
1 & 0%
\end{bmatrix}%
,  \tag{80}
\end{equation}%
and%
\begin{equation}
\Gamma _{W}^{i}\equiv 
\begin{bmatrix}
0 & \rho ^{i} \\ 
-\rho ^{i} & 0%
\end{bmatrix}%
.  \tag{81}
\end{equation}%
By promoting the variable $\hat{p}_{\mu }\rightarrow -i\partial _{\mu }$,
one recognizes in (78) the Dirac type equation in the Weyl representation in
(2+2)-dimensions. Observe that all the $\Gamma _{W}^{\mu }$ are real. So in
principle it is not necessary to look for the Majorana representation.
Nevertheless, just to see what are the differences with respect the steps of
the previous section let us now construct also the Dirac equation in the
Dirac and the Majorana representations in (2+2)-dimensions.

Adding and subtracting (75) and (76) leads to

\begin{equation}
\hat{p}_{0}(\varphi _{R}+\varphi _{L})+\sigma ^{i}\hat{p}_{i}(\varphi
_{L}-\varphi _{R})+M_{0}(\varphi _{R}+\varphi _{L})=0  \tag{82}
\end{equation}%
and%
\begin{equation}
-\hat{p}_{0}(\varphi _{L}-\varphi _{R})-\sigma ^{i}\hat{p}_{i}(\varphi
_{R}+\varphi _{L})+M_{0}(\varphi _{L}-\varphi _{R})=0,  \tag{83}
\end{equation}%
respectively. So by defining

\begin{equation}
\begin{array}{c}
\varphi _{A}=\varphi _{R}+\varphi _{L}, \\ 
\\ 
\varphi _{B}=\varphi _{L}-\varphi _{R},%
\end{array}
\tag{84}
\end{equation}%
one finds

\begin{equation}
\hat{p}_{0}\varphi _{A}+\rho ^{i}\hat{p}_{i}\varphi _{B}+M_{0}\varphi _{A}=0,
\tag{85}
\end{equation}%
and

\begin{equation}
-\hat{p}_{0}\varphi _{B}-\rho ^{i}\hat{p}_{i}\varphi _{A}+M_{0}\varphi
_{B}=0.  \tag{86}
\end{equation}%
Thus, (85) and (86) lead to

\begin{equation}
\left[ \left( 
\begin{array}{cc}
1 & 0 \\ 
0 & -1%
\end{array}%
\right) \hat{p}_{0}+\left( 
\begin{array}{cc}
0 & \rho ^{i} \\ 
-\rho ^{i} & 0%
\end{array}%
\right) \hat{p}_{i}+\left( 
\begin{array}{cc}
1 & 0 \\ 
0 & 1%
\end{array}%
\right) M_{0}\right] \left( 
\begin{array}{c}
\varphi _{A} \\ 
\varphi _{B}%
\end{array}%
\right) =0,  \tag{87}
\end{equation}%
and consequently using the definitions

\begin{equation}
\Gamma _{D}^{0}=\left( 
\begin{array}{cc}
1 & 0 \\ 
0 & -1%
\end{array}%
\right) \text{ \  \ , }\  \  \Gamma _{D}^{i}=\left( 
\begin{array}{cc}
0 & \rho ^{i} \\ 
-\rho ^{i} & 0%
\end{array}%
\right) ,  \tag{88}
\end{equation}%
the expression (87) becomes%
\begin{equation}
\left[ \Gamma _{D}^{0}\hat{p}_{0}+\Gamma _{D}^{i}\hat{p}_{i}+M_{0}\right]
\varphi _{D}=0  \tag{89}
\end{equation}%
or

\begin{equation}
\left[ \Gamma _{D}^{\mu }\hat{p}_{\mu }+M_{0}\right] \varphi _{D}=0. 
\tag{90}
\end{equation}%
Here the $D$ in $\Gamma _{D}^{\mu }$ and $\varphi _{D}$ means Dirac
representation.

Again the Weyl and the Dirac representations are related by

\begin{equation}
\left( 
\begin{array}{c}
\varphi _{A} \\ 
\varphi _{B}%
\end{array}%
\right) =\left( 
\begin{array}{cc}
1 & 1 \\ 
-1 & 1%
\end{array}%
\right) \left( 
\begin{array}{c}
\varphi _{R} \\ 
\varphi _{L}%
\end{array}%
\right) =\left( 
\begin{array}{c}
\varphi _{R}+\varphi _{L} \\ 
\varphi _{L}-\varphi _{R}%
\end{array}%
\right) .  \tag{91}
\end{equation}%
Let us denote

\begin{equation}
W=\left( 
\begin{array}{cc}
1 & 1 \\ 
-1 & 1%
\end{array}%
\right) .  \tag{92}
\end{equation}%
One finds

\begin{equation}
W^{-1}=\frac{1}{2}\left( 
\begin{array}{cc}
1 & -1 \\ 
1 & 1%
\end{array}%
\right) .  \tag{93}
\end{equation}%
One can show that $\Gamma _{W}^{\mu }$ and $\Gamma _{D}^{\mu }$ are related
by

\begin{equation}
\Gamma _{D}^{\mu }=W\Gamma _{W}^{\mu }W^{-1}.  \tag{94}
\end{equation}

Let us now expand (85) and (86) in the form

\begin{equation}
\hat{p}_{0}\varphi _{A}+\rho ^{1}\hat{p}_{1}\varphi _{B}+\rho ^{2}\hat{p}%
_{2}\varphi _{B}+\rho ^{3}\hat{p}_{3}\varphi _{B}+M_{0}\varphi _{A}=0, 
\tag{95}
\end{equation}%
and

\begin{equation}
-\hat{p}_{0}\varphi _{B}-\rho ^{1}\hat{p}_{1}\varphi _{A}-\rho ^{2}\hat{p}%
_{2}\varphi _{A}-\rho ^{3}\hat{p}_{3}\varphi _{A}+M_{0}\varphi _{B}=0, 
\tag{96}
\end{equation}%
respectively. Multiplying (95) and (96) by $i\rho ^{2}$ one obtains

\begin{equation}
\hat{p}_{0}i\rho ^{2}\varphi _{A}+i\rho ^{2}\rho ^{1}\hat{p}_{1}\varphi
_{B}+i\rho ^{2}\rho ^{2}\hat{p}_{2}\varphi _{B}+i\rho ^{2}\rho ^{3}\hat{p}%
_{3}\varphi _{B}+M_{0}i\rho ^{2}\varphi _{A}=0,  \tag{97}
\end{equation}%
and

\begin{equation}
-\hat{p}_{0}i\rho ^{2}\varphi _{B}-i\rho ^{2}\rho ^{1}\hat{p}_{1}\varphi
_{A}-i\rho ^{2}\rho ^{2}\hat{p}_{2}\varphi _{A}-i\rho ^{2}\rho ^{3}\hat{p}%
_{3}\varphi _{A}+M_{0}i\rho ^{2}\varphi _{B}=0.  \tag{98}
\end{equation}%
Since

\begin{equation}
\begin{array}{c}
\rho ^{1}\rho ^{2}=-\rho ^{3}\text{ \  \ , \  \ }\rho ^{2}\rho ^{1}=\rho ^{3},
\\ 
\\ 
\rho ^{2}\rho ^{3}=-\rho ^{1}\text{ \  \ , \  \ }\rho ^{3}\rho ^{2}=\rho ^{1},%
\end{array}
\tag{99}
\end{equation}%
by adding (95) and (98) one gets

\begin{equation}
\begin{array}{c}
-i\rho ^{2}\hat{p}_{0}(\varphi _{B}-i\rho ^{2}\varphi _{A})-i\rho ^{3}\hat{p}%
_{1}(\varphi _{A}+i\rho ^{2}\varphi _{B})+\rho ^{2}\hat{p}_{2}(\varphi
_{B}-i\rho ^{2}\varphi _{A}) \\ 
\\ 
+i\rho ^{1}\hat{p}_{3}(\varphi _{A}+i\rho ^{2}\varphi _{B})+M_{0}(\varphi
_{A}+i\rho ^{2}\varphi _{B})=0,%
\end{array}
\tag{100}
\end{equation}%
while subtracting (96) and (97) one finds

\begin{equation}
\begin{array}{c}
-i\rho ^{2}\hat{p}_{0}(\varphi _{A}+i\rho ^{2}\varphi _{B})-i\rho ^{3}\hat{p}%
_{1}(\varphi _{B}-i\rho ^{2}\varphi _{A})-\rho ^{2}\hat{p}_{2}(\varphi
_{A}+i\rho ^{2}\varphi _{B}) \\ 
\\ 
+i\rho ^{1}\hat{p}_{3}(\varphi _{B}-i\rho ^{2}\varphi _{A})+M_{0}(\varphi
_{B}-i\rho ^{2}\varphi _{A})=0.%
\end{array}
\tag{101}
\end{equation}%
Defining%
\begin{equation}
\begin{array}{c}
\varphi _{1}=\varphi _{A}+i\rho ^{2}\varphi _{B} \\ 
\\ 
\varphi _{2}=\varphi _{B}-i\rho ^{2}\varphi _{A}%
\end{array}
\tag{102}
\end{equation}%
the expressions (100) and (101) become

\begin{equation}
-i\rho ^{2}\hat{p}_{0}\varphi _{2}-i\rho ^{3}\hat{p}_{1}\varphi _{1}+\rho
^{2}\hat{p}_{2}\varphi _{2}+i\rho ^{1}\hat{p}_{3}\varphi _{1}+M\varphi
_{1}=0,  \tag{103}
\end{equation}%
and%
\begin{equation}
-\rho ^{2}\hat{p}_{0}\varphi _{1}-i\rho ^{3}\hat{p}_{1}\varphi _{2}-\rho ^{2}%
\hat{p}_{2}\varphi _{1}+i\rho ^{1}\hat{p}_{3}\varphi _{2}+M_{0}\varphi
_{2}=0.  \tag{104}
\end{equation}%
respectively. In turn, the equation (103) and (104) can be written as

\begin{equation}
\begin{array}{c}
\left[ 
\begin{array}{c}
\left( 
\begin{array}{cc}
0 & -i\rho ^{2} \\ 
-i\rho ^{2} & 0%
\end{array}%
\right) \hat{p}_{0}+\left( 
\begin{array}{cc}
-i\rho ^{3} & 0 \\ 
0 & -i\rho ^{3}%
\end{array}%
\right) \hat{p}_{1}+\left( 
\begin{array}{cc}
0 & \rho ^{2} \\ 
-\rho ^{2} & 0%
\end{array}%
\right) \hat{p}_{2} \\ 
\\ 
+\left( 
\begin{array}{cc}
i\rho ^{1} & 0 \\ 
0 & i\rho ^{1}%
\end{array}%
\right) \hat{p}_{3}+\left( 
\begin{array}{cc}
1 & 0 \\ 
0 & 1%
\end{array}%
\right) M_{0}%
\end{array}%
\right] \\ 
\\ 
\times \left( 
\begin{array}{c}
\varphi _{1} \\ 
\varphi _{2}%
\end{array}%
\right) =0.%
\end{array}
\tag{105}
\end{equation}%
Hence by using the definitions%
\begin{equation}
\begin{array}{c}
\Gamma _{M}^{0}=\left( 
\begin{array}{cc}
0 & -i\rho ^{2} \\ 
-i\rho ^{2} & 0%
\end{array}%
\right) \text{ \ , \ }\Gamma _{M}^{1}=\left( 
\begin{array}{cc}
-i\rho ^{3} & 0 \\ 
0 & -i\rho ^{3}%
\end{array}%
\right) \text{ \ , \ }\Gamma _{M}^{2}=\left( 
\begin{array}{cc}
0 & \rho ^{2} \\ 
-\rho ^{2} & 0%
\end{array}%
\right) \\ 
\\ 
\text{\  \ }\Gamma _{M}^{3}=\left( 
\begin{array}{cc}
i\rho ^{1} & 0 \\ 
0 & i\rho ^{1}%
\end{array}%
\right) \text{ \ , \ }\varphi _{M}=\left( 
\begin{array}{c}
\varphi _{A}+i\rho ^{2}\varphi _{B} \\ 
\varphi _{B}-i\rho ^{2}\varphi _{A}%
\end{array}%
\right) .%
\end{array}
\tag{106}
\end{equation}%
the equations (103) and (104) become%
\begin{equation}
\left[ \Gamma _{M}^{0}\hat{p}_{0}+\Gamma _{M}^{1}\hat{p}_{1}+\Gamma _{M}^{2}%
\hat{p}_{2}+\Gamma _{M}^{3}\hat{p}_{3}+M_{0}\right] \text{\ }\varphi _{M}=0 
\tag{107}
\end{equation}%
or

\begin{equation}
\left[ \Gamma _{M}^{\mu }\hat{p}_{\mu }+M_{0}\right] \text{\ }\varphi _{M}=0.
\tag{108}
\end{equation}%
This is Dirac equation in the Majorana representation in (2+2)-dimensions.
One can verify that

\begin{equation}
\Gamma _{M}^{\mu }\Gamma _{M}^{\nu }+\Gamma _{M}^{\nu }\Gamma _{M}^{\mu
}=-2\eta ^{\mu \nu }.  \tag{109}
\end{equation}

One can link the Dirac and the Majorana representations by the matrix

\begin{equation}
V=\left( 
\begin{array}{cc}
1 & \rho ^{2} \\ 
-\rho ^{2} & 1%
\end{array}%
\right)  \tag{110}
\end{equation}%
and its inverse%
\begin{equation}
V^{-1}=\frac{1}{2}\left( 
\begin{array}{cc}
1 & -\rho ^{2} \\ 
\rho ^{2} & 1%
\end{array}%
\right) .  \tag{111}
\end{equation}%
In fact, one has

\begin{equation}
\left( 
\begin{array}{c}
\varphi _{1} \\ 
\varphi _{2}%
\end{array}%
\right) =\left( 
\begin{array}{cc}
1 & i\rho ^{2} \\ 
-i\rho ^{2} & 1%
\end{array}%
\right) \left( 
\begin{array}{c}
\varphi _{A} \\ 
\varphi _{B}%
\end{array}%
\right) =\left( 
\begin{array}{c}
\varphi _{A}+i\rho ^{2}\varphi _{B} \\ 
\varphi _{B}-i\rho ^{2}\varphi _{A}%
\end{array}%
\right) ,  \tag{112}
\end{equation}%
and

\begin{equation}
\Gamma _{M}^{\mu }=V\Gamma _{D}^{\mu }V^{-1}.  \tag{113}
\end{equation}

It is interesting to observe that since $\Gamma _{D}^{\mu }=W\Gamma
_{W}^{\mu }W^{-1}$, one sees that

\begin{equation}
\Gamma _{M}^{\mu }=VW\Gamma _{W}^{\mu }W^{-1}V^{-1}  \tag{114}
\end{equation}%
This means that with the matrix

\begin{equation}
\begin{array}{c}
U=VW=\left( 
\begin{array}{cc}
1 & i\rho ^{2} \\ 
-i\rho ^{2} & 1%
\end{array}%
\right) \left( 
\begin{array}{cc}
1 & 1 \\ 
-1 & 1%
\end{array}%
\right) \\ 
\\ 
\text{ \  \  \  \ }=\left( 
\begin{array}{cc}
\left( 1-i\rho ^{2}\right) & \left( 1+i\rho ^{2}\right) \\ 
-\left( 1+i\rho ^{2}\right) & \left( 1-i\rho ^{2}\right)%
\end{array}%
\right) ,%
\end{array}
\tag{115}
\end{equation}%
and its inverse

\begin{equation}
\begin{array}{c}
W^{-1}V^{-1}=\frac{1}{4}\left( 
\begin{array}{cc}
1 & -1 \\ 
1 & 1%
\end{array}%
\right) \left( 
\begin{array}{cc}
1 & -i\rho ^{2} \\ 
i\rho ^{2} & 1%
\end{array}%
\right) \\ 
\\ 
\text{ \  \  \  \  \  \  \ }=\frac{1}{4}\left( 
\begin{array}{cc}
\left( 1-i\rho ^{2}\right) & -\left( 1+i\rho ^{2}\right) \\ 
\left( 1+\rho ^{2}\right) & \left( 1-i\rho ^{2}\right)%
\end{array}%
\right) ,%
\end{array}
\tag{116}
\end{equation}%
one may transform the Weyl and the Majorana $\Gamma $ matrices. In fact, one
can verify that%
\begin{equation}
\Gamma _{M}^{\mu }=\frac{1}{4}\left( 
\begin{array}{cc}
\left( 1-i\rho ^{2}\right) & \left( 1+\rho ^{2}\right) \\ 
-\left( 1+i\rho ^{2}\right) & \left( 1-i\rho ^{2}\right)%
\end{array}%
\right) \Gamma _{W}^{\mu }\left( 
\begin{array}{cc}
\left( 1-i\rho ^{2}\right) & -\left( 1+\rho ^{2}\right) \\ 
\left( 1+\rho ^{2}\right) & \left( 1-i\rho ^{2}\right)%
\end{array}%
\right) .  \tag{117}
\end{equation}

Now, we shall show that (78) is deeply linked to the $SL(2,%
\mathbb{R}
)$-group (see Ref. [19] for details). First, let us consider%
\begin{equation}
\begin{array}{cc}
\rho ^{0}=%
\begin{pmatrix}
1 & 0 \\ 
0 & 1%
\end{pmatrix}%
, & \rho ^{1}=%
\begin{pmatrix}
0 & 1 \\ 
1 & 0%
\end{pmatrix}%
, \\ 
\rho ^{2}=%
\begin{pmatrix}
0 & 1 \\ 
-1 & 0%
\end{pmatrix}%
, & \rho ^{3}=%
\begin{pmatrix}
1 & 0 \\ 
0 & -1%
\end{pmatrix}%
.%
\end{array}
\tag{118}
\end{equation}%
Notice first that the determinant of each of the matrices (118) is different
from zero. This suggests to relate such matrices with the general group $%
GL(2,%
\mathbb{R}
)$. Indeed, the matrices in (118) can be considered as a basis for a general
matrix $M$ in the following sense:

\begin{equation}
M=%
\begin{pmatrix}
A & B \\ 
C & D%
\end{pmatrix}%
=\rho ^{0}a+\rho ^{1}b+\rho ^{2}c+\rho ^{3}d,  \tag{119}
\end{equation}%
where the constants $a,b,c,d\in 
\mathbb{R}
$ are given by%
\begin{equation}
\begin{array}{cc}
a=\frac{1}{2}(A+D), & b=\frac{1}{2}(-B+C), \\ 
&  \\ 
d=\frac{1}{2}(A-D), & c=\frac{1}{2}(B+C).%
\end{array}
\tag{120}
\end{equation}%
Explicitly, (119) reads

\begin{equation}
M=%
\begin{pmatrix}
a+d & -b+c \\ 
b+c & a-d%
\end{pmatrix}%
.  \tag{121}
\end{equation}%
Without loss of generality, one may assume that $\det (M)\neq 0$. This means
that $M$ can be considered as element of $GL(2,%
\mathbb{R}
)$. If one also imposes the condition that $\det (M)=1$ then one finds the
matrix $M$ belongs to the Lie group $SL(2,%
\mathbb{R}
)$.

It is worthwhile mentioning that, by writing $\rho ^{i}$ in tensor notation%
\begin{equation}
\begin{array}{ccc}
\varepsilon _{ij}=\rho ^{2}, & \eta _{ij}=\rho ^{3}, & \lambda _{ij}=\rho
^{1},%
\end{array}
\tag{122}
\end{equation}%
one can construct a gravity model in two dimensions (see Ref [16] for
details).

Rewriting (75) and (76) respectively as follows

\begin{equation}
(\rho ^{0}\hat{p}_{0}+\rho ^{1}\hat{p}_{1}+\rho ^{2}\hat{p}_{2}+\rho ^{3}%
\hat{p}_{3})\varphi _{L}+M_{0}\varphi _{R}=0,  \tag{123}
\end{equation}%
and%
\begin{equation}
(\rho ^{0}\hat{p}_{0}-\rho ^{1}\hat{p}_{1}-\rho ^{2}\hat{p}_{2}-\rho ^{3}%
\hat{p}_{3})\varphi _{R}+M_{0}\varphi _{L}=0,  \tag{124}
\end{equation}%
one sees that both (123) and (124) have the matrix form (119). This means
that these two equations can be identified with the Lie group $SL(2,%
\mathbb{R}
)$. Indeed, taking into account (119), one notes that (123) and (124) can be
rewritten as

\begin{equation}
\begin{bmatrix}
\hat{p}_{0}+\hat{p}_{3} & -\hat{p}_{1}+\hat{p}_{2} \\ 
\hat{p}_{1}+\hat{p}_{2} & \hat{p}_{0}-\hat{p}_{3}%
\end{bmatrix}%
\varphi _{L}+M_{0}\varphi _{R}=0  \tag{125}
\end{equation}%
and%
\begin{equation}
\begin{bmatrix}
\hat{p}_{0}-\hat{p}_{3} & \hat{p}_{1}-\hat{p}_{2} \\ 
-\hat{p}_{1}-\hat{p}_{2} & \hat{p}_{0}+\hat{p}_{3}%
\end{bmatrix}%
\varphi _{R}+M_{0}\varphi _{L}=0,  \tag{126}
\end{equation}%
respectively. One observes that (125) and (126) are matrix-like momenta,
similar to the general matrix (119). Moreover, one can identify the momentum
matrices contained in the expressions (125) and (126) with the symmetry
group $SL(2,%
\mathbb{R}
)$. In fact, let us introduce a new momentum matrix%
\begin{equation}
\mathcal{P}^{\pm }=\frac{1}{M_{0}}%
\begin{bmatrix}
\hat{p}_{0}\pm \hat{p}_{3} & \pm (-\hat{p}_{1}+\hat{p}_{2}) \\ 
\pm (\hat{p}_{1}+\hat{p}_{2}) & \hat{p}_{0}\mp \hat{p}_{3}%
\end{bmatrix}%
.  \tag{127}
\end{equation}%
Consequently, the equations (125) and (126) become%
\begin{equation}
\mathcal{\hat{P}}^{+}\varphi _{L}+\varphi _{R}=0,  \tag{128}
\end{equation}%
and%
\begin{equation}
\mathcal{\hat{P}}^{-}\varphi _{R}+\varphi _{L}=0,  \tag{129}
\end{equation}%
respectively. Note that taking into account the constraint (11) we have

\begin{equation}
\det \mathcal{\hat{P}}^{\pm }\varphi _{R,L}=\varphi _{R,L}.  \tag{130}
\end{equation}%
Symbolically, we can write%
\begin{equation}
\det \mathcal{\hat{P}}^{\pm }=I.  \tag{131}
\end{equation}%
But this means that both $\mathcal{\hat{P}}^{+}$ and $\mathcal{\hat{P}}^{-}$
are elements of $SL(2,%
\mathbb{R}
)$-group and therefore the Dirac type equation (125) or (126) has a
structure associated with the group $SL(2,%
\mathbb{R}
)^{+}\times SL(2,%
\mathbb{R}
)^{-}$. In fact, this may be understood by considering the isomorphism $%
SO(2,2)\sim SL(2,%
\mathbb{R}
)\times SL(2,%
\mathbb{R}
)$.

As it is known, the Dirac equation describes massive spin-$\frac{1}{2}$
particles. When the mass $m_{0}$ is the mass of the electron, the Dirac
equation correctly determines the quantum theory of the electron. On the
other hand, the Dirac type equation (128) and (129) in (2+2)-dimensions also
describes massive spin-$\frac{1}{2}$ particles. However, there is a
significant distinction for this signature: while the Dirac equation in
(1+3)-dimensions $\psi $ can be choose as a Majorana or Weyl spinor (but not
both at the same time), in the case of (2+2)-dimensions one can choose $%
\varphi $ as a Majorana and Weyl spinor simultaneously.

\bigskip \ 

\noindent \textbf{5. Final Comments}

\smallskip \ 

\noindent It is known that $SL(2,R)$-symmetry and Lorentz symmetry $SO(t,s)$
imply together that the signatures (1+1) and (2+2) are exceptional [1].
Motivated by this result we have develop first a possible systematic method
to find the Weyl, Dirac and Majorana representations of the Dirac equation
in (1+3)-dimensions. Then we apply similar mechanism to case of the Dirac
equation in (2+2)-dimensions. Specifically, we find the Weyl, Dirac and
Majorana representations of the Dirac equation in (2+2)-dimensions. Focusing
in the gamma matrices $\gamma $, one finds that in (1+3)-dimensions and in
the Weyl representation, one has three real and one imaginary gamma matrix
and four imaginary matrices in the Majorana representation, while in
(2+2)-dimensions one has four real matrices in the Weyl representation and
three imaginary and one real in the Majorana representation: this seems to
imply some kind of duality between the signatures (1+3) and (2+2).

In (1+3)-dimensions the Majorana representation of the gamma matrices tell
us that $\gamma _{M}^{\mu }$ is pure imaginary. Therefore taking the complex
conjugate of (56),%
\begin{equation}
\left[ -\gamma _{M}^{\ast \mu }\hat{p}_{\mu }+m_{0}\right] \text{\ }\psi
_{M}^{\ast }=0,  \tag{132}
\end{equation}%
and the fact that

\begin{equation}
\gamma _{M}^{\mu }=-\gamma _{M}^{\ast \mu },  \tag{133}
\end{equation}%
one discovers that $\psi _{M}^{\ast }$ satisfies%
\begin{equation}
\left[ \gamma _{M}^{\mu }\hat{p}_{\mu }+m_{0}\right] \text{\ }\psi
_{M}^{\ast }=0,  \tag{134}
\end{equation}%
and therefore in this case it makes sense to choose

\begin{equation}
\psi _{M}^{\ast }=\psi _{M}.  \tag{135}
\end{equation}%
This is the Majorana condition. Spin-$\frac{1}{2}$ particle satisfying (134)
and (135) are called Majorana fermions. It is worthwhile mentioning that the
Majorana fermions have been proposed as fundamental structure of dark matter.

Now, the question arises what are the corresponding equations (133)-(135) in
the Weyl representation. Since $\gamma _{M}^{\mu }$ and $\gamma _{W}^{\mu }$
are related by (62) one has%
\begin{equation}
\gamma _{M}^{\ast \mu }=V^{\ast }W^{\ast }\gamma _{W}^{\ast \mu }W^{-1\ast
}V^{-1\ast }=U^{\ast }\gamma _{W}^{\ast \mu }U^{-1\ast },  \tag{136}
\end{equation}%
where we used (63). Thus, combining (62), (133) and (136) one learns that%
\begin{equation}
-\gamma _{W}^{\mu }=U^{-1}U^{\ast }\gamma _{W}^{\ast \mu }U^{-1\ast }U. 
\tag{137}
\end{equation}%
This implies that we have found a matrix

\begin{equation}
B=U^{-1}U^{\ast },  \tag{138}
\end{equation}%
which transform the Dirac equation in the Weyl representation (see eq. (26))
into the form

\begin{equation}
\left[ \gamma _{W}^{\mu }\hat{p}_{\mu }+m_{0}\right] \text{\ }\psi
_{W}^{c}=0,  \tag{139}
\end{equation}%
where%
\begin{equation}
\psi _{W}^{c}=B\psi _{W}^{\ast }.  \tag{140}
\end{equation}%
Thus the Majorana condition (135) in the Weyl representation becomes

\begin{equation}
\psi _{W}^{c}=B\psi _{W}^{\ast }=\psi _{W}.  \tag{141}
\end{equation}%
Explicitly one has

\begin{equation}
B=\left( 
\begin{array}{cc}
0 & -\sigma ^{2} \\ 
\sigma ^{2} & 0%
\end{array}%
\right) ,  \tag{142}
\end{equation}%
and therefore (141) implies

\begin{equation}
\left( 
\begin{array}{cc}
0 & -\sigma ^{2} \\ 
\sigma ^{2} & 0%
\end{array}%
\right) \left( 
\begin{array}{c}
\psi _{R}^{\ast } \\ 
\psi _{L}^{\ast }%
\end{array}%
\right) =\left( 
\begin{array}{c}
\psi _{R} \\ 
\psi _{L}%
\end{array}%
\right) ,  \tag{143}
\end{equation}%
which means that

\begin{equation}
-\sigma ^{2}\psi _{L}^{\ast }=\psi _{R},  \tag{144}
\end{equation}%
or

\begin{equation}
\sigma ^{2}\psi _{R}^{\ast }=\psi _{L}.  \tag{145}
\end{equation}%
Consequently, the Majorana spinor in the Weyl representation looks like

\begin{equation}
\left( 
\begin{array}{c}
\psi _{R} \\ 
\sigma ^{2}\psi _{R}^{\ast }%
\end{array}%
\right) .  \tag{146}
\end{equation}

In the case of (2+2)-dimensions one may start by taking the complex
conjugate to the Dirac equation in the Weyl representation given in (78),
namely%
\begin{equation}
(\Gamma _{W}^{\mu \ast }\hat{p}_{\mu }^{\ast }+M_{0})\varphi _{W}^{\ast }=0.
\tag{147}
\end{equation}%
But in (2+2)-dimensions the $\Gamma _{W}^{\mu }$ are real quantities.
Therefore the equation (147) becomes

\begin{equation}
(\Gamma _{W}^{\mu }\hat{p}_{\mu }-M_{0})\varphi _{W}^{\ast }=0.  \tag{148}
\end{equation}%
We observe that (148) differs from (78) in the sign of the mass term $M_{0}$%
, so only if $M_{0}=0$, one can set $\varphi _{W}=\varphi _{W}^{\ast }$. So,
only if the particle is massless it is possible to have a real spinor in
(2+2)-dimensions. Result that it is completely different to the case of
(1+3)-dimensions.

It is worth mentioning why we called left $\psi _{L}$ and right $\psi _{R}$
spinors in (22) and (23) for (1+3)-dimensions (and left $\varphi _{L}$ and
right $\varphi _{R}$ in (75) and (76) for (2+2)-dimensions). This is related
to the fact that in the massless case, that is when $m_{0}=0$, the spinors $%
\psi _{L}$ and $\psi _{R}$ are eigenspinors of the helicity operator $\sigma
^{j}\hat{p}$. In fact, in this case equations (22) and (23) are reduced to

\begin{equation}
(\hat{p}_{0}+\sigma ^{j}\hat{p}_{j})\psi _{L}=0  \tag{149}
\end{equation}%
and

\begin{equation}
(\hat{p}_{0}-\sigma ^{i}\hat{p}_{i})\psi _{R}=0,  \tag{150}
\end{equation}%
respectively. But, of course, in the case of $m_{0}\neq 0$ the quantities $%
\psi _{L}$ and $\psi _{R}$ are not eigenstates of such helicity operator and
in fact one finds that $\psi _{R}$ and $\psi _{R}$ are not decouple states
as in the expressions (149) and (150). In this sense, perhaps, in the case $%
m_{0}\neq 0$, it could be a better notation to write $\psi _{L}$ as $\psi
_{-}$ and $\psi _{R}$ as $\psi _{+}$ and only in the limit $m_{0}\rightarrow
0$ to assume that $\psi _{+}\rightarrow $ $\psi _{R}$ and $\psi
_{-}\rightarrow \psi _{L}$. In the case of (2+2)-dimensions the situation is
more subtle, since we do not even know whether the operator $\rho ^{j}\hat{p}%
_{j}$ is linked with the spin of the system. Just as $\sigma ^{j}$ is linked
to the representation of the group $SU(2)$, the variables $\rho ^{j}$ must
be related to the group $SL(2,R)$ which spin representation is more subtle
and requires further investigation.

In the introduction we just made some comments on the importance of the
(2+2)-signature in different scenarios of physics and mathematics. But we
still need to emphasize the possible physical relevance of such a signature.
Assume that one can make the transformations (2+2)$\leftrightarrows $(1+3).
This means that one can interchange one space dimension by one time
dimension and vice versa. It turns out that this kind of transformation has
already been done in the context of the sigma model (see Ref. [20] and
references therein). But, perhaps, still more interesting should be to
consider such a transformation in a kind of a cosmological model. One can
imagine a cosmological model in which `before' the big-bang the signature
was (2+2) and then as the universe evolves a phase transition occurred in
such a way that the signature changed from (2+2) to (1+3)-dimensions. From
the point of view of the Dirac equation this means that before the big-bang
there exist a kind of fermions in (2+2)-dimensions. But due to such a phase
transition the fermions in (1+2) changed to fermions in (1+3)-dimensions and
in particular some of these fermions could lead to electrons, neutrinos, or
Majorana fermions. In particular, for instance, in the phase transition one
may expect a broken symmetry which may lead to the correct value of the
electron mass.

Finally, it may be interesting for further research to consider the present
formalism from the point of view of supersymmetry. Since supersymmetry has
been linked to oriented matroid theory (see Refs. [21]-[22] and references
therein) one wonders whether there is a link between Majorana fermions and
oriented matroids.

\bigskip

\begin{center}
\textbf{Acknowledgments}
\end{center}

This work was partially supported by PROFAPI-UAS 2013.

\end{document}